\documentstyle[12pt]{article}

\begin{document}

\LaTeX{}

\bigskip\ \bigskip\ \bigskip\ 

\begin{center}
Q1D organic metals : the electronic contribution to the equation of state 
\bigskip\ 

\bigskip\ 

Vladan Celebonovic \smallskip\ 

Institute of Physics,Pregrevica 118,11080 Zemun-Beograd,Yugoslavia \medskip\ 

celebonovic@exp.phy.bg.ac.yu

\newpage\ 
\end{center}

Abstract: Starting from the recently derived expression for the chemical
potential of the electron gas on a 1D lattice,and a well known
thermodynamical relation,we have obtained the equation of state and the
specific heat of the electron gas on a 1D lattice .

\newpage

Introduction \medskip\ 

The knowledge of the equation of state ( EOS\ )\ is of crucial importance in
studies of a variety of physical systems. These range from the early
Universe and the quark confinement transition [1],through stellar and
planetary interiors [2],[3], to various problems encountered in ''ordinary''
solid state physics [4] .

It is the purpose of this note to determine the electronic thermal
contribution to the EOS of quasi one dimensional ( Q1D ) organic
metals.These materials were discovered in 1980.[5],[6],but according to a
recent bibliographical search on the WWW their EOS is an open problem. Q1D
organic metals are most often studied theoretically within the Hubbard model
of correlated electrons. The calculation to be reported in the next section
is motivated by recent work on the electrical conductivity of the Q1D
organic metals [7] .\medskip\ 

The calculations \medskip\ 

Neglecting electron-phonon scattering,the EOS of a solid can be expressed as
[4]

\begin{equation}
\label{(1)}P=P_c+P_{T_a}+P_{T_e} 
\end{equation}

\medskip\ 

The first term in eq.(1) denotes the pressure at T = 0 K,the second one is
the contribution of the vibrations of atoms ( or ions ) in the solid,and the
final term denotes the thermal contribution of the electron gas. The object
of this communication is to determine the third term in eq.( 1 ),using as
input the recently obtained result for the chemical potential of the
electron gas on a 1D lattice [9]. Mathematically,the calculation will be
performed using the thermodynamical relation [8,sect.24].

\begin{equation}
\label{(2)}d\mu =-sdT+vdP 
\end{equation}
\medskip\ The symbols s and v denote the enthropy and volume per particle,$%
\mu $ is the chemical potential.\newpage\ 

\medskip\ 

It follows from eq.( 2 ) that

\begin{equation}
\label{(3)}\frac{\partial \mu }{\partial T}=-\frac{\partial s}{\partial T}%
dT-s+\frac{\partial v}{\partial T}dP+v\frac{\partial P}{\partial T} 
\end{equation}

\medskip\ Assuming that s and v are temperature independent,one gets from
eq.(3) that

\begin{equation}
\label{(4)}\frac{\partial \mu }{\partial T}=-s+v\frac{\partial P}{\partial T}
\end{equation}

\medskip\ and finally

\begin{equation}
\label{(5)}\frac{\partial P}{\partial T}=\frac 1v(\frac{\partial \mu }{%
\partial T}+s) 
\end{equation}

\medskip\ 

Equation ( 5 ) is a differential form of the EOS.Obviously,in the particular
case of Q1D organic metals,the volume is equivalent to the length of the
specimen,and the volume per particle is synonymous to the lattice constant.

The chemical potential of the electron gas on a 1D lattice is [9]

\begin{equation}
\label{(6)}\mu =\frac{(\beta t)^6(na-1)\left| t\right| }{1.1029+.1694(\beta
t)^2+.0654(\beta t)^4} 
\end{equation}

\medskip\ where a denotes the lattice constant,$\beta $ is the inverse
temperature,t the hopping and n the band filling.Deriving eq.( 6 ) with
respect to T,multiplying out the products and powers,expressing the result
as a sum,one finally gets

\begin{equation}
\label{(7)}\frac{\partial \mu }{\partial T}\cong \frac{-.3388t^8\left|
t\right| }{k_B^8T^9[1.1029+.1694(\beta t)^2+.0654(\beta t)^4]^2}+<<5>> 
\end{equation}

\medskip\ where $<<5>>$ denotes the number of omitted terms.Developing
eq.(7) in its full form into series up to and including terms of the order T$%
^3$ it follows that \newpage\ 

\medskip\ 

\begin{equation}
\label{(8)}\frac{\partial \mu }{\partial T}\cong (1-an)\left| t\right|
\left[ 30.581\frac{t^2}{k_B^2T^3}+310.541(\frac{k_B}t)^2T^3-420.37(\frac{k_B}%
t)^4T^3+..\right] 
\end{equation}
$\ $

\medskip\ Inserting eq.(\ 8 ) into eq. ( 5 )\ and integrating,one gets the
following final form of the thermal contribution of the electron gas to the
EOS of Q1D organic metals

\medskip\ 
\begin{equation}
\label{(9)}P_{T_e}v=sT+(1-an)\frac{\left| t\right| }{(\beta t)^2}%
\{[155.2705-1070.0925(\beta t)^{-2}+..]-15.2905(\beta t)^4\} 
\end{equation}
$\ \qquad \qquad \qquad \qquad \qquad \qquad \qquad \qquad \qquad \qquad
\qquad \quad \ $

\medskip\ 

Discussion and conclusions\medskip\ 

Expression ( 9 ) represents the thermal EOS of the electron gas on a 1D
lattice.The existence of the lattice and of the band structure of a solid
leads to obvious differences between our result and the EOS of a free
electron gas ( such as given in,for example, [8],sect.56 ).

Equation ( 9 ) is an approximate result,because of the developement in
powers of T in eq. (\ 8 ) . Such a developement\ is physically justified
because of the fact that all studies of Q1D organic metals are performed at
low temperature.

Our result has potentially important implications.It is often stated that
Q1D organic metals have to be studied within the Luttinger liqud theory (
however,see [10] ). The specific heat of a Luttinger liqud is linear in
temperature ( for example [11] ). On the other hand,applying the relation

\begin{equation}
\label{(10)}\left( \frac{\partial c_v}{\partial v}\right) _T=T\left( \frac{%
\partial ^2P}{\partial v^2}\right) 
\end{equation}

\medskip\ ( [8] , sect. 16 ) , one gets the following expression ,\medskip\ 

\begin{equation}
\label{(11)}c_v\cong 15.2905(1-an)(\frac t{k_B})^2\frac{\left| t\right| }%
T\frac 1{v^2}-\frac{sT^2}{v^2}+o(T^3) 
\end{equation}

\medskip\ which is clearly non-linear \ in temperature.It thus follows that
by measuring the dependence of the electronic specific heat on the
temperature,it could be possible to make an experimental distinction between
the applicability of Fermi liquid and Luttinger liquid theories to Q1D
organic metals.\ 

Note that the behaviour of the pressure in eq. ( 9 ) is heavily influenced
by the band filling, which is expectable. In the special case of a half -
filled band,and taking a = 1 , the EOS reduces to the following simple form

\begin{equation}
\label{(12)}P_{T_e}v=sT 
\end{equation}

\medskip\ Obtaining a complete EOS of a Q1D organic metal would require the
determination of the first two terms in eq. ( 1 ).Such work is currently in
preparation.

\bigskip\ 

References\medskip\ 

[1] Knowles,I.G.and Lafferty,G.D.,preprint CERN-PPE/97-040 ( 1997 ).

[2] Burrows,A., Hubbard,W.B., Lunine,J.I.,Marley,M.S.et al., preprint

astro-ph / 9705027 (1997).

[3] Dappen,W.,Bull.Astr.Soc.India,{\bf 24},151 ( 1996 ).

[4] Eliezer,S.,Ghatak,A. and Hora,H.,An introduction to equations of

state: theory and applications, Cambridge Univ.Press, Cambridge

( 1986 ).

[5] Bechgaard,K., Jacobsen,C.S., Mortensen,K. et.al., Solid State

Comm.,{\bf 33},1119 ( 1980\ ).

[6] Jerome,D.,Mazaud,A.,Ribault,M.and Bechgaard,K.,J.Physique

Lettres ( Paris ),{\bf 41},L95 (1980).

[7] Celebonovic,V., Phys.Low-Dim.Struct.,{\bf 3/4},65 ( 1997 ).

[8] Landau,L.D. and Lifchitz,E.M.,Statisticheskaya Fizika,Vol.1,

Nauka,Moscow ( 1976 ).

[9] Celebonovic,V., Phys.Low-Dim.Struct.,{\bf 11/12}, 25 ( 1996 ).

[10] Bourbonnais,C.,preprint cond-mat/9611064 v2 ( 1996 ).

[11] Voit,J.,Rep.Progr.Phys.,{\bf 58},975 ( 1995 ).

\end{document}